\documentstyle[psfig,12pt]{article}
\begin{document}
\begin{titlepage}

\bigskip
\begin{center}
        {\Large {\bf Determination of CKM phases through rigid
polygons of
	flavor SU(3) amplitudes}}\\
\end{center}
\bigskip
\begin{center}
 {\large Amol S. Dighe \footnote{Email address:
		asdighe@kimbark.uchicago.edu}\\}
        Enrico Fermi Institute and Department of Physics\\
        University of Chicago, Chicago, IL 60637\\
\end{center}
\medskip
\bigskip


\begin{abstract}
\baselineskip = 24pt
Some new methods for the extraction of CKM phases $\alpha$ and
$\gamma$ using
flavor SU(3) symmetry are suggested. Rigid polygons are
constructed
in the complex plane with sides equal to the decay
amplitudes of B
mesons into two light (charmless) pseudoscalar mesons. These
rigid polygons incorporate all the possible amplitude triangles and,
being
overdetermined, also serve as consistency checks and in estimating
the rates of
some decay modes. The same techniques also lead to numerous useful
amplitude
triangles when octet-singlet mixing has been taken into account and
nearly
physical $\eta,\eta'$ are used.
\end{abstract}
\end{titlepage}


\section{Introduction}

In the standard model, CP violation is parametrized by the
Cabibbo - Kobayashi - Maskawa (CKM) \cite{ckm1} matrix. 
The unitarity
of
the CKM
matrix implies $V_{ub}^*V_{ud} + V_{cb}^*V_{cd} + V_{tb}^*V_{td} =
0$,
suggesting a unitarity triangle with its three sides as these three
$V^*V$
terms and the angles $\alpha,\beta,\gamma
\;(\alpha+\beta+\gamma=\pi)$ which,
in Wolfenstein's parametrization \cite{wolf}, take the form
\begin{equation}
\alpha={\rm Arg}(-\frac{V_{td}}{V_{ub}^*}),\;\;\beta=-{\rm
Arg}(V_{td}),\;\;
\gamma={\rm Arg}(V_{ub}^*).
\end{equation}
The current data from the measurements of
$\frac{\epsilon'}{\epsilon}$,
$B-\overline{B}$ mixing, and $\frac{|V_{ub}|}{|V_{cb}|}$
\cite{ranges} give
the allowed ranges of these three phases (95\% c.l.) as
\begin{equation}
-1.00 \leq \sin 2\alpha \leq 1.00,\;\;\; 0.21 \leq \sin 2\beta \leq
0.93,\;\;\;
0.12 \leq \sin^2\gamma \leq 1.00~~~.
\end{equation}
The decays of B mesons to light pseudoscalar mesons (B $\to$ PP) give
us
access to the third row and third column of the CKM matrix, where all
the above
phases lie. Experiments will give us
the magnitudes of the amplitudes of the decay into
various decay
channels. (The data about the time-dependence of the decays will be
available,
but we shall not use that here). If theory expects the amplitudes of
some three
decays to form a triangle in the complex plane, constructing this
triangle from
the experimentally measured amplitudes will give us the relative
phases between
these amplitudes, from which information about
the three phases above can be obtained.

Assuming flavour SU(3) symmetry gives us numerous such triangle
relations and
ways to determine these phases. Some such ways, with or without using
any
time-dependent information, have been suggested in
\cite{zeppen}-\cite{buras} and the extent of SU(3) breaking
effects has been estimated \cite{ghlrsu3} to be about 20\%.

The major contributions to the amplitudes of decays are from the tree
or
penguin type diagrams. The tree diagrams involve the process $b \to
uW$ with
the other quark in the B meson acting as a spectator, whereas
penguins are
taken to be dominated by $t$-quark exchange.
(Corrections to the
$t$-quark dominance of the $b \to d$ and $b \to s$ QCD penguin
amplitudes \cite{fleischer}
have been neglected in this analysis.)
The phases contributed to various
types of diagrams by the CKM matrix elements in the dominating term
are as
shown in Table~(\ref{phases}).

\begin{table}
\label{phases}
\begin{center}
\begin{tabular}{|l|c|c|}
\hline
Diagram type&	$|\Delta S|=0$&			$|\Delta S|=1$\\
\hline
Tree&		${\rm Arg}(V_{ub}^*V_{ud})=\gamma$& ${\rm
Arg}(V_{ub}^*V_{us})=\gamma$\\
\hline
Penguin& ${\rm Arg}(V_{tb}^*V_{td})=-\beta$& ${\rm
Arg}(V_{tb}^*V_{ts})=0$\\
\hline
\end{tabular}
\end{center}
\caption{The weak phases for tree and penguin.}
\end{table}

Section~\ref{rep} discusses the representation of
decay amplitudes
in terms of an SU(3)
invariant basis. Section~\ref{polygons} gives the
``rigid polygon'' relations between
these amplitudes. Sections~\ref{s=1} and ~\ref{s=0} outline
the strategies for extracting CKM phases
from the relative phases of the amplitudes for $|\Delta S| = 1$
and $\Delta S = 0$ respectively. They also illustrate these
methods with examples of particular decay modes and comment
on their experimental feasibility.
Section~\ref{eta} discusses the singlet-octet $\eta$ mixing and
the additional amplitudes
introduced due to this. Section~\ref{triang} gives some of
the corresponding useful
amplitude
triangles  when approximate physical particles
$\eta,\eta'$ are used instead of $\eta_8,\eta_1$.
Section~\ref{conclude} concludes.


\section{Representation of amplitudes within SU(3)}
\label{rep}

Two main approaches have been taken \cite{ghlrtr,ghlrd50,dh2,savage}
for
finding the amplitude triangle (or quadrilateral) relations. One
method is
to represent the amplitudes in terms of the basis of $T$ (tree), $P$
(penguin),
 $C$ (color-suppressed tree), $E$ (exchange), $A$ (annihilation) and
$PA$ (penguin
annihilation) diagram contributions. An exhaustive list of all such
amplitudes
has been made in \cite{ghlrd50} and some quadrangle, triangle or
equivalence
relations have been obained. The contributions by $E$, $A$, $PA$
have been neglected
since they are expected to be suppressed
by a factor of $\frac{f_B}{m_B}=5\%$.
($E$ and
$A$ will also be helicity suppressed by a factor $\frac{m_q}{m_b}$
where
$q=u,d,s$).

Another equivalent approach \cite{dh2,savage} is to represent
the amplitudes in the basis
of six SU(3) invariant amplitudes whose combinatorial coefficients
will be the
6 invariant quantities formed by the combinations of the 3-vector
$B_i
\equiv(B^+,B^0,B_s)$, two pseudoscalar matrices $M^i_j$ (one for each
pseudoscalar), and $H$, the hamiltonian for $\overline{b} \to
\overline{q_1} q_2 \overline{q_3}$. $H$ can be split using SU(3) into
$\overline{3} \otimes 3 \otimes \overline{3} =
\overline{3} \oplus \overline{3} \oplus 6 \oplus \overline{15}$
and thus its
transformation
properties can be encoded into $H^i(3)$, a vector, $H^{[ij]}_k(6)$, a
traceless tensor
antisymmetric in the upper two indices and $H^{(ij)}_k(15)$, a
traceless tensor
symmetric
in the upper two indices.

The tree amplitude, in terms of this basis, will be (modulo the CKM
factors)
\begin{eqnarray}
T = 	&   A_3^T B_i H^i(3) M^j_k M^k_j &
	\mbox{+}  C_3^T B_i M^i_j M^j_k H^k(3) \nonumber \\
 \mbox{}	& + A_6^T B_i H^{ij}_k(6) M^l_j M^k_l &
	\mbox{+}  C_6^T B_i M^i_j H^{jk}_l(6) M^l_k \nonumber \\
 \mbox{}	& + A_{15}^T B_i H^{ij}_k(15) M^l_j M^k_l &
	\mbox{+}  C_{15}^T B_i M^i_j H^{jk}_l(15) M^l_k~~~.
\end{eqnarray}
The $A_i^T$ are the terms that come from contracting the light quark
part of
the B-vector directly with the hamiltonian. This would imply that the
light
quark is an active part of the decay process and not just a
spectator. Such
amplitudes (corresponding to E, A, PA)  will
be
suppressed by a factor of $\frac{f_B}{m_B}$ and hence can be
neglected to a
first approximation. The tests for this approximation (which will
come
naturally from the method of grouping described below) are specified
at the end of this section.

The values of the nonzero elements of $H(3),H(6),H(15)$ are given in
\cite{dh2,savage}. The convention used for the members of the
pseudoscalar meson
octet is
\begin{eqnarray}
\pi^+=u\overline{d}, &
\pi^0=\frac{1}{\sqrt{2}}(d\overline{d}-u\overline{u}), &
\pi^-=-\overline{u}d,  \nonumber \\ K^+=u\overline{s}, &
K^0=d\overline{s}, &
\overline{K^0}=\overline{d}s,
 \nonumber \\
 K^-=-\overline{u}s, &
\eta_8=\frac{1}{\sqrt{6}} (2s\overline{s}- u\overline{u}-
d\overline{d}).
\end{eqnarray}
An exhaustive list of all the coefficients in $B \to PP$ is given in
Table~\ref{su3coeff}.

\begin{table}
\label{su3coeff}
\begin{tabular}{|l|lll||rrr|rrr|r|l|}
\hline
Group & \multicolumn{3}{c||}{Decay mode} &
\multicolumn{6}{c|}{Coefficients of}
& Factor & Linear \\
no. & & & & $C_3$ & $C_6$ & $C_{15}$ & $A_3$ & $A_6$ & $A_{15}$ & &
Combination
\\
\hline
$1$ &
$B^+$ &          $K^+$ &  $\overline{K^0}$ &         1 &    $-1$ &
$-1$
&     0 &
   1 &     3 & 1 & $p$ \\
& $B^0$ &          $K^0$ &  $\overline{K^0}$ &        1 &    $-1$ &
 $-1$
&     2 &
  1 &    $-3$ & 1 & \\
& $B^0$ &         $\pi^0$ &        $\eta_8$ &         1 &    $-1$ &
 $-1$
&     0
&     1 &   $-5$ & $-\sqrt{3}$ & \\
\hline
$1'$ &
$B_s$ &          $K^0$ &  $\overline{K^0}$ &         1 &    $-1$ &
$-1$
&     2 &
   1 &    $-3$ & 1 & $p$ \\
 & $B^+$ &          $K^0$ &         $\pi^+$ &         1 &    $-1$ &
 $-1$
&     0
&     1 &     3 & 1 & \\
\hline
$2'$ & $B^+$ &          $K^+$ &         $\pi^0$ &         1 &    $-1$
&     7
&
  0 &     1 &     3 & $-\sqrt{2}$ & $t+c+p$ \\
\hline
$3$ & $B^+$ &         $\pi^+$ &         $\pi^0$ &         0 &     0 &
    8 &
  0 &     0 &     0 & $-\sqrt{2}$ & $t+c$ \\
\hline
$4'$ & $B^+$ &          $K^+$ &        $\eta_8$ &        1 &    $-1$
&    $-9$
 &
 0 &     1 &     3 & $\sqrt{6}$ & $-t-c+p$ \\
\hline
$5$ & $B^+$ &         $\pi^+$ &        $\eta_8$ &        $-2 $&     2
&    $-6$
 &
  0 &    -2 &    -6 & $\sqrt{6}$ & $-t-c-2p$ \\
\hline
$6$ & $B^0$ &         $\pi^0$ &         $\pi^0$ &         1 &     1 &
  $-5$
&
 2 &    $-1$ &     1 & $\sqrt{2}$ & $-c+p$ \\
 & $B_s$ &  $\overline{K^0}$ &         $\pi^0$ &        1 &    1 &
 $-5$
&     0
 &     $-1$ &     $-1$ & $\sqrt{2}$ & \\
 & $B_s$ &  $\overline{K^0}$ &        $\eta_8$ &         1 &     1 &
  $-5$
&
0 &    $-1$ &    $-1$ & $\sqrt{6}$ & \\
\hline
$6'$ & $B^0$ &          $K^0$ &         $\pi^0$ &        1 &    1 &
  $-5$
&
 0 &     $-1$ &     $-1$ & $\sqrt{2}$  & $-c+p$ \\
 & $B^0$ &          $K^0$ &        $\eta_8$ &         1 &     1 &
$-5$
&     0
&    $-1$ &    $-1$ & $\sqrt{6}$ & \\
\hline
$7'$ & $B_s$ &         $\pi^0$ &        $\eta_8$ &         0 &
$-2$ &     4
 &
   0 &     2 &    $-4$ & $-\sqrt{2}$ &  $c$ \\
\hline
$8$ & $B^0$ &         $\pi^+$ &         $\pi^-$ &         1 &     1 &
    3 &
  2 &    $-1$ &     1 & $-1$ & $t+p$ \\
 & $B_s$ &          $K^-$ &         $\pi^+$ &        1 &    1 &    3
&     0 &
   $-1$ &     $-1$ & $-1$ & \\
\hline
$8'$ & $B^0$ &          $K^+$ &         $\pi^-$ &         1 &     1 &
    3 &
  0 &    $-1$ &    $-1$ & $-1$ & $t+p$ \\
 & $B_s$ &          $K^+$ &          $K^-$ &        1 &    1 &    3 &
   2
& $-1$
 &    1 & $-1$ & \\
\hline
$9$ & $B^0$ &        $\eta_8$ &        $\eta_8$ &         1 &    $-3$
&  3 &
 6 &     3 &    $-3$ & 3$\sqrt{2}$ & $c+p$ \\
\hline
$10$ & $B^0$ &          $K^+$ &          $K^-$ &         0 &     0 &
   0 &
$-2$ &     0 &    $-2$ & 1 & 0 \\
\hline
$10'$ & $B_s$ &         $\pi^+$ &         $\pi^-$ &         0 &     0
&     0 &
    2 &     0 &     2 & $-1$ & 0 \\
 & $B_s$ &         $\pi^0$ &         $\pi^0$ &         0 &     0 &
 0 &
4 &     0 &     4 & $-1$ & \\
\hline
$11'$ & $B_s$ &        $\eta_8$ &        $\eta_8$ &         2 &     0
&   $-6$
 &
  3 &     0 &   $-3$ & 3$/\sqrt{2}$ & $-c+2p$ \\
\hline
\end{tabular}
\caption{The coefficients of the six invariant amplitudes under
SU(3).}
\end{table}

The decay modes having the same $C_i$ coefficients are grouped
together and
given the same group number. The unprimed (primed) group numbers
correspond to
the decay modes with $|\Delta S|=$ 0 (1). $A(j)$ represents the
amplitude for
all the decay modes in that group. The advantage of using the
amplitudes in
terms of group numbers is that in case of multiple decay modes in a
group, the
ones easier to detect can be used or a suitable average of all the
modes within
a group can be taken to improve statistics.
Comparisons of branching fractions of decay modes within a group
also serve as tests of flavor SU(3) symmetry.

The {\em factor} column gives the factor by which the actual decay
amplitude
should be multiplied to get the {\em group} amplitude, e.g.
\begin{equation}
        A^T(4') \equiv  \sqrt{6} A^T(B^+ \to K^+ \eta_8)~~~.
\end{equation}
The {\em linear combination} column gives the group amplitudes
in terms of three
complex numbers $t,c,p$; where we neglect the $A_i^T$ terms. This
helps in
getting the triangle relations, and will be referred to in
Section~\ref{eta}
where the number of distinct amplitudes is greater and drawing
polygon
relations is not particularly instructive.

The penguin part of the amplitude can similarly be written in terms
of
$A_i^P$'s and $C_i^P$'s. The coefficients of
$A_i^T(j),A_i^T(j'),A_i^P(j)$, and
$A_i^P(j')$ have the same value and so do the coefficients of
$C_i^T(j),C_i^T(j'),C_i^P(j)$, and $C_i^P(j')$.
Henceforth, the superscripts $P$ and $T$ on $A_i$ and $C_i$
will be omitted wherever the
relations hold true for both types of amplitudes.

The 12 amplitudes $A_i^T,\ A_i^P,\ C_i^T$,\ and $C_i^P$
($i = 3,6,15$) are not all
independent, since there can be only 5 independent amplitudes.
[When combined with the triplet light quark in the B meson,
$ \overline{3} \otimes 3 = 1 \oplus 8_1 $,
$ 6 \otimes 3 = 8_2 \oplus 10 $,
$ \overline{15} \otimes 3 = 8_3 \oplus \overline{10} \oplus 27 $.
{}From PP $( 8 \otimes 8 = 1 \oplus 8 \oplus 8 \oplus 10
\oplus \overline{10} \oplus 27)$, we get three symmetric states,
one singlet, one (symmetrized) octet and one 27-plet.
The coupling of these two implies that the decays are
characterised by one singlet, three octets and one 27-plet,
a total of 5 independent amplitudes.] With $A_i$ terms
neglected, the number of independent amplitudes reduces to three.
There are 3 relations between the six amplitudes $C_i^{T(P)}$,
given by
\begin{eqnarray}
C_3^T = C_6^T + C_{15}^T & ,&  C_6^P = C_{15}^P = 0~~.
\end{eqnarray}

The net amplitudes for the decay modes can be written as
\begin{eqnarray}
\label{PTcomb1}
A(j) & = A^P(j)+A^T(j) & = V_{tb}^*V_{td}P(j) + V_{ub}^*V_{ud}T(j) \\
\label{PTcomb2}
A(j') & = A^P(j')+A^T(j') & = V_{tb}^*V_{ts}P(j') +
V_{ub}^*V_{us}T(j')
\end{eqnarray}
where $P=|P|e^{i\delta_P}$ and $T=|T|e^{i\delta_T}$ are the
penguin- and tree-
type contributions modulo the CKM factors, and $A^P$ and $A^T$
include the CKM
factors.

We clearly have the relations
\begin{eqnarray}
\label{ppprime}
A^P(j')=\frac{V_{ts}}{V_{td}} A^P(j) & , &
A^T(j')=\frac{V_{us}}{V_{ud}} A^T(j)
\end{eqnarray}
between the primed and the unprimed amplitudes, but nothing can be
inferred
{\it a priori} about the relationship between $A(j)$ and $A(j')$
unless either
one of $A^P$ or $A^T$ can be neglected.

Electroweak penguins do not have the same  SU(3) representations
as QCD penguins (the $u$ and $d$
quarks
definitely interact with the photon with different strengths).
But they always appear with the $T$, $C$, and $P$ diagrams
in
fixed combinations (the left hand sides of
equations in ~(\ref{tcp})),
so the triangle relations based solely on the basis of matching of
$T$, $C$, and $P$ diagrams hold true even in the presence of these
electroweak
penguins.

The {\it linear combination} column can be translated in the language
of
$T,\ C,\ P,\ P_{EW}$ diagrams \cite{ghlrewpeng} as (the superscript
{\em C}
 stands for {\em color-suppressed})
\begin{eqnarray}
\label{tcp}
t=T + (c_u-c_d) P_{EW}^C~, &
c=C + (c_u - c_d) P_{EW}~, &
p=P + c_d P_{EW}^C~,
\end{eqnarray}
where $c_u,\ c_d,\ (c_s=c_d)$ are the strengths with which u, d, and
s quarks,
respectively, interact with $Z$ and $\gamma$ in electroweak penguins.

Neglecting $A_i^T,A_i^P$ contributions gives $A(10)=A(10')=0$. This
thus serves
as a test for the validity of this approximation. The rates of the
decays $B^0
\to K^+K^-,B_s \to \pi^+\pi^-,B_s \to \pi^0\pi^0$ should be
suppressed by
$\frac{f_B}{m_B}$.


\section{The ``rigid polygon'' relations}
\label{polygons}

\begin{figure}
\psfig{file=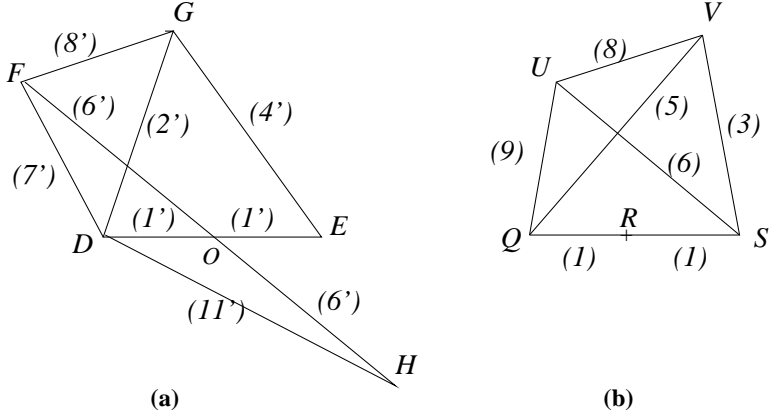,width=4in}
\caption{The rigid amplitude polygons (a) $|\Delta S|=1$ (b) $\Delta
S =0$ .}
\label{poly}
\end{figure}

An amplitude triangle is formed by three decay modes $a,b,c$
iff there exist three numbers $ n_a, n_b, n_c $ such that
$n_a A(a) + n_b A(b) + n_c A(c) = 0$, i.e.
\begin{equation}
n_a C_i(a) + n_b C_i(b) + n_c C_i(c) = 0 \quad {\mbox for}
\quad i=3,6,15~~~,
\end{equation}
where $C_i(a)$ is the coefficient of $C_i$ in the decay mode $a$.
We shall denote a triangle formed with the sides $A(a),A(b),A(c)$ as
$\bigtriangleup(a-b-c)$ where $a,b,c$ are the corresponding group
numbers. All
possible triangles with the $B \to PP$ decay amplitudes as their
sides can be
found and represented concisely in the form of two distinct rigid
polygons, one
each for $|\Delta S|=1$ and $\Delta S=0$. (See Figure~\ref{poly}.)
The polygons are overdetermined, so the amplitudes and phases of some
of the
decay modes can be predicted from the others.

The polygons are oriented such that the penguin-only decay modes are
along the
real axis. These are the modes with $A^T=0$ so that their net phase
can be
written simply as $\delta_P+{\rm Arg}(V_{tb}^*V_{tq})$ where $q =
(d,~s)$ for
$|\Delta S| = (0,~1)$. When we draw these rigid polygons, we thus
know the
phases of all the amplitudes modulo this phase, since all the decay
modes are
connected via these rigid polygons to either $(1)$ or $(1')$.

One major advantage of having penguin-only decay modes in the
polygons is that
for these decays, $|A(1)|=|\tilde{A}(1)|$ and
$|A(1')|=|\tilde{A}(1')|$ where
$\tilde{A}(j)$ is the amplitude of the CP-conjugate process in which
all the
particles in $(j)$ have been replaced by their antiparticles. While
superposing
the ``particle'' and ``antiparticle'' triangles in the process of
extraction of
the CKM phases, we shall always superpose their penguin-only sides.


\section{CKM phases from $|\Delta S| = 1$ triangles}
\label{s=1}

In this case, the penguin-only modes for both particle and
antiparticle decay
have the phase $\delta_P+\pi$, so that aligning these amplitudes
along the real
axis is equivalent to rotating the amplitude triangles by an unknown
but fixed
phase, $\delta_P$.

As a result of this rotation, the amplitude of a generic decay is
now, from
Eq.~(\ref{PTcomb2}),
\begin{equation}
A_R(i')=-|V_{tb}^*V_{ts}||P|+|V_{ub}^*V_{us}||T|
	e^{i(\delta_T-\delta_P)}e^{i\gamma},
\end{equation}
where the subscript R stands for the amplitudes with their phase
rotated such
that the phase of the penguin-only amplitude is $0$ or $\pi$.

The amplitude for the antiparticle decay is
\begin{equation}
\tilde{A}_R(i')=-|V_{tb}^*V_{ts}||P|+|V_{ub}^*V_{us}||T|
		e^{i(\delta_T-\delta_P)}e^{-i\gamma}.
\end{equation}
Subtracting the two equations gets rid of the penguin contribution.
\begin{eqnarray}
A_R(i')-\tilde{A}_R(i') & = & 2i \sin\gamma
|V_{ub}^*V_{us}||T|e^{i(\delta_T-\delta_P)} \nonumber \\
 & = & 2i \sin\gamma |A^T(i')| e^{i(\delta_T-\delta_P)}~~~.
\label{singamma}
\end{eqnarray}
Now, if the tree contributions to $|A^T(i')|$ and $|A^T(j)|$ can be
related for
some $j$ and furthermore if the penguin contribution to $A(j)$ can be
neglected
(as is the case whenever the tree contribution is not
color-suppressed
\cite{ghlrewpeng}), then the measurement of the decay rate gives us
directly
the value of $|A^T(i)|$, hence $|A^T(i')|$, and from
Eq.~(\ref{singamma}) we
obtain the value of sin $\gamma$ and $\delta_T-\delta_P$.

\subsection{\underline{$\bigtriangleup(1'-2'-4')$}}
\label{first}

\begin{equation}
\label{oneone}
A(4')=2A(1')-A(2')~~~.
\end{equation}
Constructing the CP-conjugate triangle with
$\tilde{A}(4'),\tilde{A}(2'),
\tilde{A}(1')$ and superposing $A(1')$ and $\tilde{A}(1')$ (this is
possible
since $|A(1')|=|\tilde{A}(1')|$; this will introduce a discrete
twofold
ambiguity, leading to a discrete twofold ambiguity in sin $\gamma$),
the line
joining the remaining vertices of these two triangles is
\begin{equation}
A_R(2')-\tilde{A}_R(2')=2i \sin\gamma
|A^T(2')|e^{i(\delta_T-\delta_P)}~~~.
\label{onetwo}
\end{equation}
Also,
\begin{equation}
|A^T(2')|= |A^T_{K^+\pi^0}|= \frac{|V_{us}|}{|V_{ud}|}
\frac{f_K}{f_\pi}
|A^T_{\pi^+ \pi^0}|= \frac{|V_{us}|}{|V_{ud}|} \frac{f_K}{f_\pi}
		|A^T(3)|~~~.
\end{equation}
The factor of $f_K / f_{\pi}$ comes from taking into account
the first order SU(3) breaking under the assumption
of factorization.
The
dominating contribution to $|A^T_{K^+ \pi^0}| (|A^T_{\pi^+ \pi^0}|)$
is from the tree diagram, in which $K^+ (\pi^+)$ is formed
purely through a weak current $[W^+ \to K^+ (\pi^+)]$. With
factorization, this
implies a multiplicative
term  $<K^+|\overline{s} \gamma_{\mu} \gamma_5 u|0>
(<\pi^+|\overline{d} \gamma_{\mu} \gamma_5 u|0>) $
in the amplitude, which is proportional to $f_K (f_{\pi})$.
This is precisely where the first order SU(3) breaking appears
\cite{ghlrsu3}.

$B^+ \to \pi^+ \pi^0$ proceeds only via $I=2$ channel, so the QCD
penguin does not contribute here. (QCD penguin is a pure
$\Delta I = 1/2$ operator.) Since the electroweak penguin can
be neglected in $\Delta S = 0$ channels \cite{ghlrewpeng},
$|A^T(3)| \approx |A(3)|$
and the measurement of $|A(3)|$ along with
Eq.~(\ref{onetwo}) gives the value of sin $\gamma$
and
$\delta_T-\delta_P$ up to a twofold ambiguity.

If factorization holds here to a fair extent, then
\begin{equation}
|A_R(2')-\tilde{A}_R(2')| \leq 2 |A^T(2')| \approx
\frac{|V_{us}|}{|V_{ud}|} \frac{2 f_K}{f_{\pi}} |A(3)|.
\end{equation}
This provides a weak partial test (necessary, but not sufficient)
for factorization.

The same triangle has been suggested by Deshpande and He in
\cite{dh2}. It is
restated here for the sake of completeness and to illustrate the
method. The
decay modes involved here have at least one charged particle in the
final
states (in group $(1')$, we can choose $B^+ \to K^0 \pi^+$) and the
branching
fractions are expected to be ${\cal O}(10^{-5})$, so this triangle
will be
experimentally easy to construct. The difficulty of separating
$\eta_8$ haunts
it, though (as it does all the modes in this section that involve
$\eta_8$).

\subsection{\underline{$\bigtriangleup(2'-7'-8')$}}
\label{second}

We already know the phase of $A_R(2')$ from the construction of
$\bigtriangleup(1'-2'-4')$.
Now we can use
\begin{equation}
A(2')=A(7')+A(8')
\end{equation}
and construct this triangle on top of the earlier one. Constructing
the
CP-conjugate triangle and orienting it using the information about
the phase
of $\tilde{A}_R(2')$ from the construction in Sec.~\ref{first}, we
get
the phases of $A_R(8')$ and $\tilde{A}_R(8')$. Now
\begin{equation}
A_R(8')-\tilde{A}_R(8')=2i \sin\gamma |A^T(8')|
e^{i(\delta_T-\delta_P)}
\end{equation}
and assuming factorization,
\begin{equation}
|A^T(8')|=|A^T_{B^0\to K^+\pi^-}|
	=\frac{|V_{us}|}{|V_{ud}|} \frac{f_K}{f_\pi} |A^T_{B^0 \to \pi^+
\pi^-}|
	=\frac{|V_{us}|}{|V_{ud}|}\frac{f_K}{f_\pi}|A^T(8)|~~.
\end{equation}
If we can neglect the QCD penguin contribution in $A(8)$, then
$|A^T(8)| \approx |A(8)|$ and we get the value of sin
$\gamma$ and
$\delta_T-\delta_P$ up to a discrete twofold ambiguity. The argument
about the factor $f_K/f_{\pi}$ in Sec.~\ref{first}
and a similar weak partial test for factorization
\begin{equation}
|A_R(8')-\tilde{A}_R(8')| \leq \frac{|V_{us}|}{|V_{ud}|}
	\frac{2 f_K}{f_{\pi}} |A(8)|
\end{equation}
is valid in this case also.

Alternatively,, knowing $\gamma$ and $A_R(2'),\tilde{A}_R(2')$
from
Sec.~\ref{first}, the measurement of $|A(7')|,|\tilde{A}(7')|$ will
enable us
to determine $A_R(8'), \tilde{A}_R(8')$ simply by geometry.

This triangle has been suggested in \cite{ghlrewpeng} as a part of
the
quadrilateral FGOD (Fig~\ref{poly}) with a slightly different
approach.
Both the decay products in the mode $(7')$ are neutral particles. So
we might
come across the problem of low acceptance rate here. The branching
fraction
for this mode is also expected to be very small \cite{dh3}.

\section{CKM phases from $\Delta S=0$ triangles}
\label{s=0}

When the polygon is oriented such that the penguin-only amplitude is
along the
real axis, the amplitude [Eq.~(\ref{PTcomb1})] of a generic decay
becomes
\begin{equation}
A_R(j)=|V_{tb}^*V_{td}||P(j)|+
|V_{ub}^*V_{ud}||T(j)|e^{i(\delta_T-\delta_P)}
				e^{i(\pi-\alpha)}
\end{equation}
since ${\rm Arg}(\frac{V_{ub}^*V_{ud}}{V_{tb}^*V_{td}})=\pi-\alpha$.

The corresponding antiparticle amplitude oriented in a similar manner
will be
\begin{equation}
\tilde{A}_R(j)=|V_{tb}^*V_{td}||P(j)|+ |V_{ub}^*V_{ud}||T(j)|
			e^{i(\delta_T-\delta_P)}e^{i(-\pi+\alpha)}~~~.
\end{equation}
When the tree contribution in $\overline{b} \to
\overline{u}u\overline{d}$ is
not color suppressed, the penguin contribution is expected to be
$\sim \lambda
(\approx 0.2)$ times the tree contribution \cite{ghlrewpeng} and
the $|P(j)|$ term can
be neglected. The angle between these two amplitudes will then be
$2\alpha$.

\subsection{\underline{$\bigtriangleup(1-5-3)$}}
\label{third}
\begin{equation}
A(3)=-A(5)-2A(1)
\end{equation}
and superposing the CP-conjugate triangle such that $A_R(1)$ and
$\tilde{A}_R(1)$
overlap, we get (up to a discrete twofold ambiguity) the relative
phases of
$A_R(5)$ and $\tilde{A}_R(5)$, which is $2\alpha$.

$|A(1)|$ is expected to be very small and one might have to worry
about low
statistics here. But we know that $A(1)=A^P(1),A(1')=A^P(1')$ and
from
Eq.~(\ref{ppprime}), we get
\begin{equation}
A(1)=\frac{V_{td}}{V_{ts}}A(1').
\end{equation}
If the value of $\frac{|V_{td}|}{|V_{ts}|}$ is known through other
means, we
can use the data from $A(1')$ to determine the magnitude of $A(1)$ to
be used
in the construction of this triangle. $(3)$ and $(5)$ have one
charged particle
in their final state and a $T$ contribution (in the $\Delta S = 0$
mode) which
would indicate a sizeable branching fraction and higher acceptance
for both of
them.

\subsection{\underline{$\bigtriangleup(1-9-6)$}}
\label{fourth}
\begin{equation}
A(9)=2A(1)-A(6)~~~.
\end{equation}
This, with its CP-conjugate triangle, gives the phases of $A_R(6)$,
$\tilde{A}_R(6)$, $A_R(9)$ and $\tilde{A}_R(9)$.

This triangle relation is not directly useful for finding any of the
CKM phases
since the tree contribution to $A(6)$ and $A(9)$
is color-suppressed and hence of the same
order of magnitude as the penguin. The penguin contribution here,
therefore,
cannot be neglected. But the information about the phases
will be used in the construction of the next triangle.

\subsection{\underline{$\bigtriangleup(9-5-8)$ or
$\bigtriangleup(6-3-8)$}}
\label{fifth}
\begin{equation}
A(9)+A(8)=-A(5)
\end{equation}
or
\begin{equation}
A(8)-A(6)=A(3),
\label{iso}
\end{equation}
with the information about the phase of $A_R(6)$ or $A_R(9)$ from
Sec.~\ref{fourth}, gives the phase of $A_R(8)$ and
the CP-conjugate triangle gives
the phase of $\tilde{A}_R(8)$. The phase difference between these two
is
$2\alpha$, similar to Sec.~\ref{third}.

$\bigtriangleup(6-3-8)$ is the same as the $\pi-\pi$ isospin triangle
in
\cite{isospin}.
With $(3),(5),(8)$ having T contributions (and consequently, a
charged
particle in the final state), the branching fractions and the
acceptance for
these modes is expected to be on the higher side.

\medskip

$\bigtriangleup(1'-6'-7')$ -- a part of the quadrilateral suggested
in
\cite{ghlrewpeng}
-- and $\bigtriangleup(1'-6'-11')$ are possible, but are not very
useful since
the tree
contribution to $A(7')$, $A(6')$ and $A(11')$ is color-suppressed and
the
electroweak penguin contributions is expected to be significant.


\section{Physical $\eta$ and $\eta'$}
\label{eta}

The SU(3) eigenstates $\eta_8,\eta_1$ are different from the physical
particles
$\eta,\eta'$. Taking into account the mixing angle of $\approx
20^\circ$
\cite{mixing}, these physical states are very close to
\cite{ghlrd50,BR}
\begin{equation}
\eta=\frac{1}{\sqrt{3}} (s\overline{s}- u\overline{u}-
d\overline{d})~~~,
\label{et}
\end{equation}
\begin{equation}
\eta'=\frac{1}{\sqrt{6}} (2s\overline{s}+ u\overline{u}+
d\overline{d})~~~.
\label{et'}
\end{equation}

Since we have contributions from the singlet component here, the
trace of
the pseudoscalar matrix M is no longer zero. There will, therefore,
be
additional terms in the amplitude, whose coefficients would have been
zero
had we been dealing with only the pseudoscalar octet mesons.

The additional terms in the amplitude will be
\begin{eqnarray}
  E_3 B_i M^i_j H^j(3) M^k_k & + &  D_3 B_i H^i(3) M^j_j M^k_k
\nonumber \\
	& + &   D_6 B_i H^{ij}_k(6) M^k_j M^l_l \nonumber \\
	& + & 	 D_{15} B_i H^{ij}_k(15) M^k_j M^l_l ~~~.
\label{addl}
\end{eqnarray}
There are no terms with amplitudes $E_6$ or $E_{15}$ since their
coefficients would involve $H^{ij}_j(6)$ and $H^{ij}_{j}(15)$
respectively,
all of which are zero.

The terms $D_i$ will be suppressed by $\frac{f_B}{m_B}$ since these
terms
will correspond to some annihilation diagrams (for the same reason as
for
the $A_i$'s). So the only significant additional term we have here is
the
$E_3$ term.

The coefficients of $C_i$ and $E_3$ for decays involving the physical
states $\eta,\eta'$ are given in Tables~(\ref{physprime})
and (\ref{physunprime}).


\begin{table}
\begin{tabular}{|lll||rrrr||r|l|}
\hline
\multicolumn{3}{|c||}{Decay mode} & \multicolumn{4}{c||}{Coefficients
of}
& Factor & Linear \\
 & & & $C_3$ & $C_6$ & $C_{15}$ & $E_3$ &  & combination \\
\hline
 $B^+ $&          $K^+$ &          $\eta$ &        0 &     1 &     7
&    1
& $-\sqrt{3}$ &$t+c+s$ \\
\hline
 $B^+ $&          $K^+$ &         $\eta'$ &        3 &     1 &     1
&    4
& $\sqrt{6}$ & $t+c+3p+4s$ \\
\hline
 $B_s $&         $\pi^0$ &          $\eta$ &       0 &     2 &
$-4$ &     0
& $\sqrt{6}$ &  $-c$ \\
 $B_s $&         $\pi^0$ &         $\eta'$ &        0 &    2 &
$-4$ &     0
& $\sqrt{3}$ & \\
\hline
 $B^0 $&          $K^0$ &          $\eta$ &       0 &    $-1$ &     3
&    1
& $-\sqrt{3}$ & $c+s$ \\
\hline
 $B^0 $&          $K^0$ &         $\eta'$ &        3 &    $-1$ &
$-3$ &     4
& $\sqrt{6}$ & $c+3p+4s$ \\
\hline
 $B_s $&          $\eta$ &          $\eta$ &        1 &     0 &
$-4$ & $-1$
& 3/$\sqrt{2}$ & $-c+p-s$ \\
\hline
 $B_s $&          $\eta$ &         $\eta'$ &       $-2$ &     0 &
5 & $-1$
& $-3$/$\sqrt{2}$ & $\frac{1}{2}c-2p-s$ \\
\hline
 $B_s $&         $\eta'$ &         $\eta'$ &        2 &     0 &
$-2$ &   4
& 3/$\sqrt{2}$ & $c+2p+4s$ \\
\hline
\end{tabular}
\caption{Coefficients with physical $\eta,\eta'$ for $|\Delta S|=1$
.}
\label{physprime}
\end{table}


\begin{table}
\begin{tabular}{|lll||rrrr||r|l|}
\hline
\multicolumn{3}{|c||}{Decay mode} & \multicolumn{4}{c||}{Coefficients
of}
& Factor & Linear \\
 & & & $C_3$ & $C_6$ & $C_{15}$ & $E_3$  &  & combination \\
\hline
 $B^+ $&         $\pi^+$ &          $\eta$ &        2 &    $-1$ &
5 &   1
& $-\sqrt{3}$ & $t+c+2p+s$ \\
\hline
 $B^+ $&         $\pi^+$ &         $\eta'$ &        2 &     2 &     2
&   4
& $\sqrt{6}$ & $t+c+2p+4s$ \\
\hline
 $B^0 $&         $\pi^0$ &          $\eta$ &       $-2$ &     1 &
3 & $-1$
& $\sqrt{6}$ &$-2p-s$ \\
\hline
 $B^0 $&         $\pi^0$ &         $\eta'$ &       $-1 $&  $-1$ &  3
& $-2$
& $-\sqrt{3}$ & $-p-2s$ \\
\hline
 $B_s $&  $\overline{K^0}$ &          $\eta$ &        0 &    $-1$ &
  3 &  1
& $-\sqrt{3}$ & $c+s $ \\
\hline
 $B_s $&  $\overline{K^0}$ &         $\eta'$ &        3 &    $-1$ &
$-3$ &  4
& $\sqrt{6}$ & $c+3p+4s$ \\
\hline
 $B^0$ &          $\eta$ &          $\eta$ &        1 &    $-2$ &
2 &   1
& 3/$\sqrt{2}$ & $c+p+s$ \\
\hline
 $B^0 $&          $\eta$ &         $\eta'$ &        2 &    $-1$ &
1 &   5
& $-3\sqrt{2}$ & $2c+2p+5s$ \\
\hline
 $B^0 $&         $\eta'$ &         $\eta'$ &        1 &     1 &
$-1$ &   4
& 3$\sqrt{2}$ & $c+p+4s$ \\
\hline
\end{tabular}
\caption{Coefficients with physical $\eta,\eta'$ for $\Delta S=0$ .}
\label{physunprime}
\end{table}


All the amplitudes can be explicitly written in terms of four complex
numbers
$t,c,p,s$ as in the {\it linear combination} column of
Tables~(\ref{su3coeff})-(\ref{physprime}). (``$s$'' is
the additional
amplitude due to the contribution of the singlet. It is the same
as the contribution from the two-gluon diagram $P_1$ \cite{dgr}
and the corresponding electroweak penguin, $c_d P_{EW}$.)
Thus, the knowledge of the
magnitudes (obtained from the measurements) and the relative phases
(obtained
from the construction of triangles) of $t,c,p,s$ (or any four of
their
independent linear combinations) will give us the amplitudes of all
the decay
modes.
We, therefore, need only three independent connected triangles 
(each triangle shares a side with with at least one of the others)
to get the amplitudes and phases of all the other
amplitudes with the
same $|\Delta S|$.

Some examples of useful triangle relations are given in
Sec.~\ref{triang}.
 All the triangles have
discrete twofold ambiguities associated with them.

Methods for estimating the first order SU(3) breaking effects are
indicated in \cite{ghlrsu3}. One way to guage the effects of
SU(3) breaking on the amplitude triangles is to check if the
triangle relations remain valid even when the first order SU(3)
breaking terms are introduced \cite{dgr}. Phase space effects have to
be taken into account especially when the final state particles
contain one or more heavy $\eta'$.


\section{Amplitude triangles with $\eta$ and $\eta'$}
\label{triang}

\subsection{\underline{$|\Delta S|=1$}}

The triangles
\begin{eqnarray}
1. &  A(B^+ \to K^0 \pi^+)+ \sqrt{6}A(B_s \to \pi^0 \eta)
	& = \sqrt{2}A(B^0\to K^0\pi^0)~~~,
\label{phy1} \\
2. & A(B^+ \to K^0 \pi^+)+\sqrt{3}A(B^0 \to K^0 \eta)
	& = \frac{3}{\sqrt{2}}A(B_s \to \eta \eta)~~~,
\label{phy2} \\
3. & -\sqrt{3}A(B^+ \to K^+ \eta)+\frac{3}{\sqrt{2}}A(B_s \to \eta
\eta)
	& = -A(B^0 \to K^+ \pi^-)
\label{phy3}
\end{eqnarray}
are three connected amplitude triangles,
sufficient to predict the amplitudes and phases of
all the remaining decays of the type $|\Delta S|=1$.

The triangle in Eq.~(\ref{phy3}), when
constructed on the top of the triangle in Eq.~(\ref{phy2}), gives the
phase of
$A_R(B^0 \to K^+ \pi^-)$. The CP-conjugate
$A_R(\overline{B^0} \to K^- \pi^+)$ and the method of
Sec.~\ref{second} gives
$\gamma$ and $\delta_T-\delta_P$.

All the amplitudes above have penguin contributions which are
substantial
in this $|\Delta S|=1$ mode, so the branching ratios will be high,
but the
presence of many neutral particles in the final state might pose
acceptance problems.

Some more triangles, e.g.
\begin{eqnarray}
4. & A(B^+ \to K^0 \pi^+) + \frac{3}{\sqrt{2}}A(B_s \to \eta' \eta')
	& = \sqrt{6}A(B^0 \to K^0 \eta') ~~~,
\label{phy4} \\
5. & 2A(B_s \to \eta \eta) + A(B_s \to \eta' \eta')
	& = 2A(B_s \to \eta \eta')
\label{phy5}
\end{eqnarray}
can be constructed which will serve to validate our assumptions
collectively,
if not individually. The remaining amplitudes can be generated from
the
information gained through Eqs.~(\ref{phy1})-(\ref{phy5}).


\subsection{\underline{$\Delta S=0$}}

The triangle
\begin{equation}
1. \quad 3A(B^+ \to K^+ \overline{K^0})+2\sqrt{6}A(B^0 \to \pi^0
\eta)=
-\sqrt{3}A(B^0 \to \pi^0 \eta')
\end{equation}
has a penguin-only side and hence will be useful in
defining the orientations of all the other amplitudes.
\begin{equation}
2. \quad -\sqrt{2}A(B^+ \to \pi^+ \pi^0)+\sqrt{3}A(B^+ \to \pi^+
\eta)=
\sqrt{6}A(B^0 \to \pi^0 \eta)
\end{equation}
constructed
on top of the above triangle
gives the phase of $A_R(B^+ \to \pi^+ \pi^0)$. The CP-conjugate
triangle gives the phase of $A_R(B^- \to \pi^- \pi^0)$ and the phase
difference
between these is $2\alpha$. The
same
procedure can be used with the information obtained from the phases
of $A_R(B^+
\to \pi^+ \eta)$ and $A_R(B^- \to \pi^- \eta)$, the phase difference
between
which is $2\alpha$.

Both $B^+ \to \pi^+ \pi^0$ and $B^+ \to \pi^+ \eta$ have a $T$
contribution and
hence, are expected to have sizeable branching fractions and
acceptances.

These triangles, along with the $\pi-\pi$ isospin triangle
\begin{equation}
3. \quad -\sqrt{2}A(B^+ \to \pi^+ \pi^0) + \sqrt{2}A(B^0 \to \pi^0
\pi^0) =
-A(B^0 \to \pi^+ \pi^-)
\label{pipi}
\end{equation}
will enable us to predict the amplitudes and phases of all the other
decay
modes with $\Delta S=0$. The isospin triangle also enables one to get
the
phases of $A_R(B^0 \to \pi^+ \pi^-)$ and $A_R(\overline{B^0} \to
\pi^+ \pi^-)$,
the phase difference between which should be $2\alpha$.
\begin{equation}
4. \quad A(B^+ \to K^+ \overline{K^0})-\sqrt{3}A(B_s \to
\overline{K^0} \eta)=
\frac{3}{\sqrt{2}}A(B^0 \to \eta \eta)
\end{equation}
gives the phase of $A_R(B_s \to \overline{K^0} \eta)$, which is
$\delta_T-\delta_P+\gamma$ since the only contribution here is from
the $C$
diagram. (The $P_{EW}$ contribution is expected to be $\sim \lambda
(\approx 0.2)$ times the $C$ contribution here \cite{ghlrewpeng}).
The CP-conjugate triangle gives the phase of $A_R(\overline{B_s} \to
K^0 \eta)$, $\delta_T-\delta_P-\gamma$ and thus, $\gamma$ is obtained
along
with $\delta_T-\delta_P$. (This is one instance where
$\delta_T-\delta_P$ is
obtained in the $\Delta S=0$ mode. But this one will be plagued by
low
statistics and more neutral particles in the final state.)

Some additional triangles like
\begin{eqnarray}
5. & 2A(B^+ \to K^+ \overline{K^0}) + 3\sqrt{2}A(B^0 \to \eta' \eta')
	& =
	\sqrt{6}A(B_s \to \overline{K^0} \eta')~~~ ,\\
6. & A(B^0 \to \eta \eta) + 2A(B^0 \to \eta' \eta')
	& = -2A(B^0 \to \eta
\eta')
\end{eqnarray}
can be constructed for consistency checks.
All the remaining amplitudes may be constructed using the information
gained from the above triangles and the {\em linear combination}
column
of Table~({\ref{physunprime}).


\section{Conclusions}
\label{conclude}

Using only the time-independent information about the rates of $B$
mesons
decaying into light pseudoscalars, we can determine the angles of the
CKM
unitarity triangle under flavor SU(3) symmetry. Here we neglect the
annihilation - type diagrams which are expected to be suppressed
by $\frac{f_B}{m_B}$. The amplitudes are represented in terms of
an SU(3) invariant basis.
Rigid amplitude polygons are constructed which are overdetermined
and hence
can serve either for multiple ways of determining $\alpha$ and
$\gamma$, as
consistency checks, as tests for the approximations made, or to
estimate the
amplitudes for decays hard to detect experimentally.
The tests for the assumptions of flavor SU(3) symmetry,
factorization, annihilation diagram suppression are also built in.

The expected branching fractions of most of the decay modes are
${\cal
O}(10^{-6}-10^{-5})$ \cite{BR} and within reach of current
and upcoming experiments. The method of grouping helps in improving
statistics by using the information   from more than one decay mode
or by
allowing one to measure, say, a mode with charged decay products
instead of
neutral ones. The knowledge of the ratios of magnitudes of CKM
elements can be
used to estimate the decay rates of some modes with lower branching
fractions.

The physical particles $\eta,\eta'$ are different from the SU(3)
singlet
$\eta_1$ or octet $\eta_8$. Taking into account this mixing, the same
methods
have been applied to the approximately physical $\eta,\eta'$, which
will be the
actual particles to be detected. The
decay modes with $\eta$ or $\eta'$ as one of the decay products form
a sizeable
portion of charmless $B$ decays and hence
taking into account the deviation of the physical states from the
octet or
singlet states is important. All the decay amplitudes to
the approximately physical particles are expressed explicitly in
terms of
four SU(3) invariant quantities and
amplitude triangle relations are found which are
directly
useful to obtain the CKM phases, validate our assumptions
and provide self-consistency tests.


\section*{Acknowledgements}

I would like to thank Aaron Grant and Mihir Worah for helpful
discussions. I am
particularly indebted to Jonathan Rosner for getting me interested in
this
topic, critically reading the manuscript and giving
helpful
suggestions. This work was supported in part by the U. S. Department
of Energy
under Contract No. DE FG02 90ER40560.


\end{document}